\documentstyle[psfig]{l-aa}
\def\zabs{$z_{\rm abs}$}
\def\zem{$z_{\rm em}$~}
\def\lya{Ly$\alpha$ }
\def\lyb{Ly$\beta$ }
\def\ovi{O~{\sc vi}~ }

\def\nv{N~{\sc v}~ }

\def\ciii{C~{\sc iii}$\lambda$977~ }

\def\civ{C~{\sc iv}~}

\def\nva{N~{\sc v}$\lambda$1238~ }
\def\nvb{N~{\sc v}$\lambda$1242~ }

\def\kms{km~s$^{-1}$}

\begin{document}
\thesaurus{11.17.1;11.17.4 J2233-606}
\title{
The $z_{\rm abs}$~$\sim$~$z_{\rm em}$ absorption line systems 
toward QSO J2233-606 in the Hubble Deep Field South:
Ne~{\sc viii}$\lambda\lambda$770,780 absorption and partial coverage
\thanks{Based in part on
observations obtained with the NASA/ESA {\sl Hubble Space Telescope} 
by the Space Telescope Science Institute, which is operated by AURA, 
Inc., under NASA contract NAS 5--26555; observations
collected at the European Southern Observatory, La Silla, Chile; and
observations collected at the Anglo-Australian Observatory.}
}
\author{Patrick Petitjean\inst{1,2} \and R. Srianand$^3$}
\institute{$^1$Institut d'Astrophysique de Paris -- CNRS, 98bis Boulevard 
Arago, F-75014 Paris, France\\
$^2$UA CNRS 173 -- DAEC, Observatoire de Paris-Meudon, F-92195 Meudon
Cedex, France \\
$^3$IUCAA, Post Bag 4, Ganesh Khind, Pune 411 007, India }
\date{ }
\offprints{Patrick Petitjean}
\maketitle
\markboth{}{}
\begin{abstract}

Results of a careful analysis of the highly ionized absorption systems,
observed over the redshift range 2.198--2.2215 in the $z_{\rm
em}$~=~2.24 HDFS-QSO J2233-606, are presented. The strength and
covering factor of the \ovi and Ne~{\sc viii} absorption lines suggest
that the gas is closely associated with the AGN. In addition, most of the 
lines show signature of partial coverage and the covering factor varies from
species to species. This can be understood if the clouds cover the
continuum emission region completely and only a fraction of the broad
emission line region.


Using photo-ionization models we analyze in more detail 
the component at \zabs = 2.198, for which we can derive reliable estimates 
of column densities for H~{\sc i} and other species.
Absolute abundances are close to solar but
the [N/C] abundance ratio is larger than solar.
This result, which is consistent with the
analysis of
high-$z$ QSO broad emission-lines, confirms the physical 
association of the absorbing gas with the AGN. 
The observed column densities of N~{\sc iv}, \nv
and Ne~{\sc viii} favor a two-zone model for the absorbing region
where Ne~{\sc viii} is predominantly produced in the highly
ionized zone. 
It is most likely that in QSO J2233-606,
the region producing the Ne~{\sc viii} absorption can not be a
warm absorber. 

One of the Ly$\alpha$ absorption lines at $z_{\rm abs}$~=~2.2215
has a flat bottom typical of 
saturated lines and non-zero residual intensity in the core, consistent with
partial coverage.
There is no metal-line from this Ly$\alpha$ cloud detectable in the spectrum
which suggests either large chemical inhomogeneities in the
gas or that the gas is very highly ionized. If the latter is true 
the cloud could have a total hydrogen column density consistent with that of 
X-ray absorbers. It is therefore of first importance to check whether or not
there is an X-ray warm-absorber in front of this QSO. 



\end{abstract}

\section {Introduction}
The QSO J2233-606 ($z_{\rm em}$~=~2.24) has been given tremendous
interest as it is located in the middle of the STIS Hubble deep field
south making this field an ideal target for studying the connection
between the diffuse gaseous component of the universe and galaxies
(Ferguson 1998). The spectrum of this QSO shows
several associated systems, (i.e. systems
with \zabs~$\simeq$~\zem), at $z_{\rm abs}$~$\sim$~2.2 with broad
C~{\sc iv} and N~{\sc v} absorption lines (Sealey et al. 1998, Savaglio
1998, Outram et al.  1998). HST STIS spectra, together with the
available ground-based data, provide different pieces of information
about these systems over the rest-wavelength range 375-2800~\AA.

Associated systems have been intensively studied in the past few years
because 
they are believed to be
intimately related to the central engines of AGNs as: (i) they frequently
show absorption due to high-ionization lines; (ii) they have been convincingly
shown to have metallicities of the order of or above solar (Petitjean
et al. 1994, Savaglio et al.  1994, Hamann 1997); (iii) the absorbing gas
does not cover the background emitting region completely (Petitjean et
al. 1994, Hamann et al. 1997b, Barlow \& Sargent 1997).
In a few cases it has been shown that the optical depth of the 
high-ionization lines varies with time on scales of a year 
(Hamann et al. 1997b) as is the case for some broad absorption line systems 
(e.g. Barlow et al. 1992). This requires a recombination time-scale of less 
than a year and hence high particle density in the absorbing 
gas if the observed variability is caused by the change
in the ionizing conditions.
 

Soft X-ray spectra of an appreciable fraction of Seyfert-I galaxies show
K-shell absorption edges of ionized oxygen (O~{\sc vii} and O~{\sc
viii}; Reynolds 1997, George et al. 1998). Rapid variability of these
absorption edges suggest that the absorbing gas is very close to the
central engine. Moreover, there is a one-to-one correspondence between
the presence of associated absorption systems and
X-ray "warm absorbers" in Seyfert galaxies (Crenshaw et al.  1998).
The case for a unified model for X-ray and UV absorbers, although
attractive, is not completely convincing yet however. 
Eventhough Mathur et al. (1994)
showed that the X-ray and UV absorptions seen in 3C351 can be
reproduced by a single-cloud model, there are cases where two
different ionized zones are needed even to produce the optical depth
ratios of O~{\sc vii} and O~{\sc viii} (see Reynolds 1997).  

Some Seyfert-I galaxies show signatures of the existence of
optically thin emitting clouds in the BLR. Shield, Ferland \& Peterson
(1995) have shown that the properties of the optically thin clouds in
the inner BLR are consistent with that of warm absorbers (see also
Porquet et al. 1998). Analysis of the broad Ne~{\sc viii}$\lambda$774
{\sl emission} line in QSOs shows that the Ne~{\sc viii}-emitting
regions have ionization parameters in the range 5--30, total hydrogen
column densities of the order of 10$^{22.5}$~cm$^{-2}$ and average
covering factors $>$30\% for solar abundances and a nominal QSO
spectrum (Hamann et al. 1997b). The ionization conditions in these
emitting clouds are similar to those of warm absorbers.

In order to investigate in more detail the nature of associated systems
and their possible connection to warm absorbers, absorptions from
species with a wide range of excitation should be studied. Morevover 
column densities should be determined taking into account the effect of
partial coverage. 
In this prospect, absorption from Ne~{\sc viii}, if present, is 
crucial as its ionization potential, 207~eV, is much higher
than the ionization potential of other easily observable species.
To our knowledge the Ne~{\sc viii}$\lambda\lambda$770,780 absorption
doublet has been detected in only two associated systems, in the line
of sight to HS1700+6416 at $z_{\rm abs}$~=~2.7126 (Petitjean et al. 1996;
see the HST spectrum in Vogel \& Reimers 1995) 
and in the line of sight to UM675 (Hamann et al. 1995, see the spectrum
in Hamann et al. 1997a) at $z_{\rm abs}$~=~2.1340.
Analysis of the latter system leads the authors to conclude that, although 
the detection
of Ne~{\sc viii} provides strong evidence for a link between the
associated system and the warm absorber, the total hydrogen column
density in the Ne~{\sc viii} phase is too small to produce the warm
absorber phenomenon. Recently Telfer et al. (1998) have done a similar 
study (including detection of Ne~{\sc viii}) of the broad absorption 
line system present in QSO SBS 1542+531. Note that in all  
previous studies Ne~{\sc viii} absorption is investigated with low
dispersion FOS spectra.
%

In the following we discuss in detail the associated systems in the line of
sight to QSO J2233-606 and especially the unambiguous detection of 
strong Ne~{\sc viii} absorption. We briefly present the
method to derive column densities in case of partial coverage in Section~2;
describe the data in Section~3 and the individual absorption systems 
in Section~4, discuss 
the physical consequences of the observations in Section~5 and
draw our conclusions in Section~6.
\section {Partial coverage}
When an absorbing cloud does not cover the background source completely, the 
observed residual intensity in the normalized spectrum, $R_\lambda$, can be 
written as,
\begin{equation}
R_\lambda~=~(1-f_c)+f_c\times exp(-\tau_\lambda),
\end{equation}
where, $\tau_\lambda$ and $f_c$ are the optical depth and covering factor 
respectively. The latter is the ratio of the number of photons produced by 
the region of the background source that is occulted by the absorbing
cloud to the total number of photons (Srianand \& Shankaranarayanan 1999). 
If two absorption
lines with rest-wavelengths $\lambda_1$ and $\lambda_2$ originate from 
the same ion (usually it will be doublets or multiplets), their residual 
intensities, $R_1$ and $R_2$,  at any velocity $v$ with respect to the 
centroid of the lines are related by,
\begin{equation}
R_2(v) = 1-f_{c2}+f_{c2}\times\bigg({R_1(v)-1+f_{c1}\over f_{c1}}\bigg)^\gamma
\end{equation}
where $f_{c1}$ and $f_{c2}$ are the covering factors calculated for the two 
lines and,
$$
\gamma =\bigg({f_2\lambda_2\over f_1\lambda_1}\bigg)
$$
with $f_1$ and $f_2$ the oscillator strengths (see e.g. Petitjean 1999). 
The value of $\gamma$ is close to 2 for doublets.  \par\noindent
Though our intuition says that $f_{c1}=f_{c2}$
it need not be true in general (Srianand \& Shankaranarayanan 1999).
Indeed due the complex velocity structure in the BLR, photons that
could be absorbed by lines 1 and 2 (even in the case of doublets)
originate from spatially distinct regions in the BLR. A cloud can thus be black
in line 1 (covering the BLR region emitting at the corresponding
wavelength $\lambda_1$), but not in line 2 if it does not cover at the same 
time the region emitting photons with wavelength $\lambda_2$. 
Also, the interpretation of the relative covering factors for various
ions is not straightforward. 
Indeed, the covering factor of absorption lines 
seen over the wavelength range of the QSO spectrum that is
dominated by the continuum may be larger than the covering factor of
lines present on top of  the QSO emission lines 
as the extension of the BLR is larger than that of the continuum 
emission region.
%
\section {Data}
\begin{table}
\begin{tabular}{lcrcc}
\multicolumn{5}{l}{{\bf Table 1.} Spectroscopic data}\\ 
\hline
\multicolumn{1}{c}{Inst./Tel.}&\multicolumn{1}{c}{Coverage}&
\multicolumn{1}{c}{$R$}&\multicolumn{1}{c}{S/N}&
\multicolumn{1}{c}{Ref.}\\
\multicolumn{1}{c}{ }&\multicolumn{1}{c}{(\AA)}&
\multicolumn{1}{c}{ }&\multicolumn{1}{c}{per pixel}&
\multicolumn{1}{c}{ }\\
\hline\\
E230M-STIS/HST & 2275-3118 & 30000      & 1.5 - 4 & $^a$\\
G140L-STIS/HST & 1126-1721 & $\sim$1100 & 10 - 18 & $^a$\\
G230L-STIS/HST & 1585-3173 & $\sim$550  & 20 - 45 & $^a$\\
G430M-STIS/HST & 3025-3565 & $\sim$6000 & 5 - 15  & $^a$\\
UCLES/AAT      & 3530-4390 & 35000      & 5 - 30  & $^b$\\
EMMI/NTT       & 4386-8270 & 21000      & 8 - 10& $^c$\\
 & & & & \\
\hline 
\multicolumn{5}{l}{$^a$ Ferguson (1998), $^b$ Outram et al. (1998), $^c$
Savaglio (1998)}    
\end{tabular}
\end{table}
In the following analysis we use spectra of QSO J2233-606 obtained
with different instruments and telescopes. Table~1 gives 
the instrumentation, wavelength coverage, spectral resolution, typical
signal-to-noise ratio and reference for the data available.
Ultra-violet and optical spectra
were taken by the HST HDF-S STIS Team (Ferguson 1998) with the
STIS E230M, G140L, G230L and G430M gratings on board of the 
Hubble Space Telescope. Optical data was obtained at the AAT and 
ESO/NTT by Outram et al. (1998) and Savaglio (1998) respectively.
We have corrected the data for fluctuations in the zero-level 
by fitting low order polynomials to the bottom of the strong saturated 
Ly$\alpha$ lines.
The echelle data have been smoothed using a gaussian filter of width 3 pixels.
The continuum was fit with low-order polynomials and the
spectrum was normalized.
\section {Description of the associated system}
\subsection{Overview}
\begin{figure}
\centerline{\vbox{
\psfig{figure=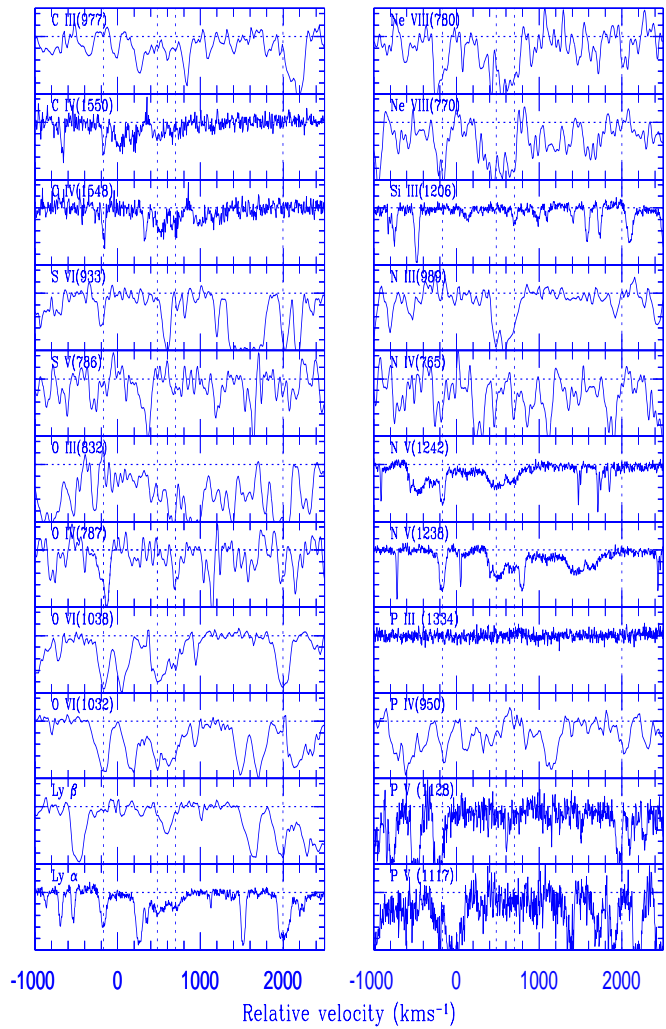,height=15.cm,width=10.cm,angle=270}
}}
\caption[]{Absorption profiles 
of different transitions 
in the associated systems observed along the line of sight to J2233-606. 
The zero velocity is taken at $z$ = 2.20. The vertical 
dashed lines mark redshifts 2.1982, 2.2052, 2.2075 and 2.2215
from the left to the right. Note the component at 
+950~km~s$^{-1}$ ($z_{\rm abs}$~=~2.21) with O~{\sc vi} and Ne~{\sc viii}
but no detectable H~{\sc i} absorptions.}
\label{transi}
\end{figure}

\begin{figure}
\centerline{\vbox{
\psfig{figure=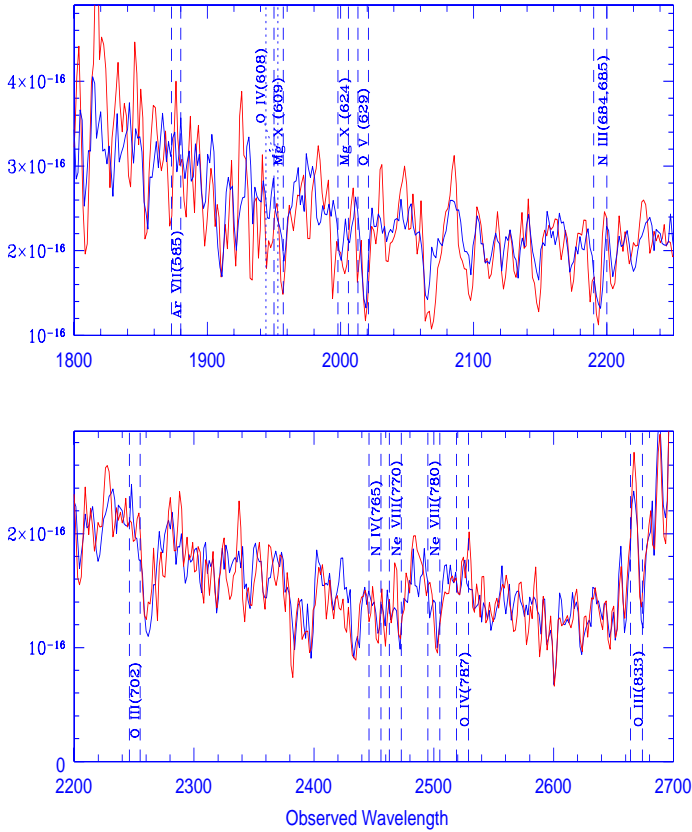,height=12.cm,width=9.5cm,angle=0}
}}
\caption[]{Possible identifications of lines from the associated system
in the G230L spectrum. The vertical 
dashed lines mark the redshift range 2.198--2.210.
The two spectra obtained in 1997 (end of october) and 1998 
(beginning of october) are superimposed to look for variability. It can be
seen that the modest S/N ratio prevents any firm conclusion about
the variability of the 
Ne~{\sc viii} absorption lines.
}
\label{transi2}
\end{figure}

The absorption profiles produced by the associated absorbers are
shown for the most important transitions on a velocity scale 
in Fig.~\ref{transi}. Complementary identifications in the 
G230L spectrum of lower quality because of the presence of the LLS 
break at $\sim$2700~\AA~ are given in Fig.~\ref{transi2}.
The two spectra obtained in 1997 (end of october) and 1998 
(beginning of october) are superimposed on Fig.~\ref{transi2}
to look for variability. It can be
seen that the modest signal-to-noize ratio prevents any firm conclusion about
the variability of the Ne~{\sc viii} absorption lines.
There are absorption features near the expected position of
Mg~{\sc x}$\lambda\lambda$609,624 but, due to poor spectral resolution, 
this cannot be ascertained.
On the contrary, O~{\sc v} and probably N~{\sc iii} are present.
The maximum column density of He~{\sc i} found in the models 
discussed below is $\sim$10$^{11}$~cm$^{-2}$.
We thus do not believe that the possible line at $\sim$1875~\AA~ seen in
only one of the spectra can be He~{\sc i}$\lambda$584.

The emission redshift of QSO J2233-606, derived from the high-ionization 
emission lines C~{\sc iv}, C~{\sc iii}] + Al~{\sc iii}, $z_{\rm em}$~=~2.237, 
is smaller than the redshift derived from Mg~{\sc ii}, $z_{\rm em}$~=~2.252, 
by about 1390~km~s$^{-1}$ (Sealey et al. 1998). Considering the
Mg~{\sc ii} redshift as more representative of the intrinsic redshift
(Carswell et al. 1991), 
the associated absorptions, seen over 
the redshift range 2.198--2.2215, have outflow velocities
relative to the quasar of 2800--5000~km~s$^{-1}$ which is modest
compared to usual associated or BAL outflows.
\subsection{\zabs = 2.2215 }
\begin{figure}
\psfig{figure=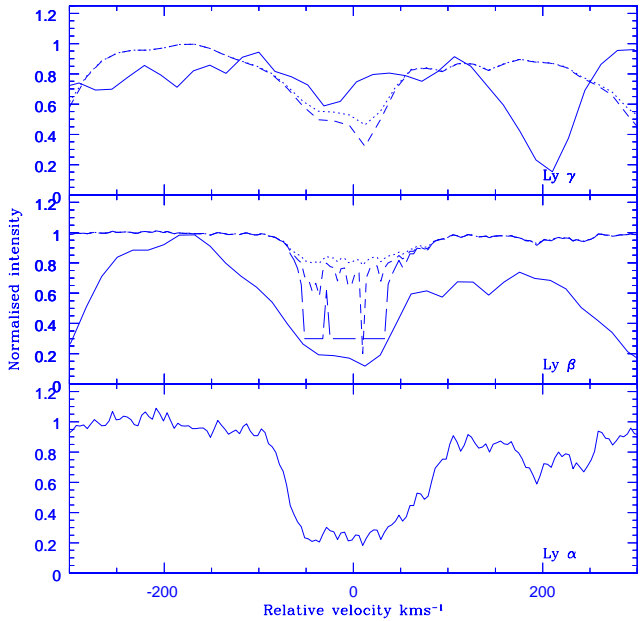,height=9.cm,angle=0}
\caption[]{Analysis of partial coverage in the \zabs = 2.2215 system.
The observed \lya profile is plotted in the bottom panel. The middle
panel shows the observed \lyb profile (solid) together with the predicted 
profiles computed from \lya assuming 
covering factors $f_{\rm c}$~=~1, 0.8 and 0.70 for the dotted, 
short-dashed and long-dashed lines respectively. 
The top panel shows the observed Ly$\gamma$ profile with the 
predicted profiles computed from \lyb assuming covering 
factors $f_{\rm c}$~=~1 and 0.9 for the dotted and dashed lines
respectively.}
\label{f222}
\end{figure}

There is a \lya line at this redshift with flat bottom, consitent with
line saturation, but with non-zero residual intensity.
From the latter, it can be seen on Fig.~\ref{f222} that the minimum
covering factor for this line is $f_{\rm c}$~$\sim$~0.7.
However before drawing any conclusions it is important to show that 
the feature is 
not due to blending of a few weaker lines.  
In Fig.~\ref{f222} we plot the line profiles of the other 
detected Lyman series lines from this system. It can be seen that 
the \lyb line is very strong. Moreover the residual intensity in the \lyb 
line is {\sl smaller} than the residual intensity in the \lya line, 
consistent with saturation of the \lya line.\par\noindent
We first assume that the covering factor is the same for \lya and \lyb.
In the middle panel we plot the
observed \lyb profile together with the predicted \lyb profiles 
computed from the \lya profile for three 
values of the covering factor: dotted, short--dashed and
long--dashed lines are for covering factors 1, 0.8, 0.70
respectively. 
%
%
\par\noindent
The predicted \lyb profiles are inconsistent with the 
observed \lyb profile and the latter 
seems too strong even for the minimum covering factor acceptable for
the \lya line ($f_{\rm c}$~$\sim$~0.7). 
%
The numerous saturated lines present in this part of the spectrum 
assures that the error in the zero level determination cannot explain
the discrepancy. Although we cannot reject the presence of weak Ly$\alpha$
absorption lines superimposed with the \lyb absorption, especially in
the blue-wing,
the good wavelength coincidence between \lyb and \lya 
seems to indicate that the contamination cannot be large.
One way to explain the apparent strength of the \lyb line is
to assume that the covering factor 
for \lyb is {\sl larger} than for \lya. Note that this is consitent
with the QSO \lya emission line to be stronger than the 
\lyb emission line. If true, then we can expect that the covering factor
of the Ly$\gamma$ line be even larger.\par\noindent
In the top panel we plot the observed wavelength range of the
Ly$\gamma$ line and the predicted profile using the \lyb profile for
covering factors 1 (dotted line) and 0.9 (dashed lines). The best match is
obtained for complete coverage though this does not reproduce the
Ly$\gamma$ very well. Smaller values of the covering factor predict
too strong a Ly$\gamma$ line. 
We conclude that, in order to understand the residual intensities
in the different Lyman series absorption lines, it must be assumed that
the covering factor increases from \lya to Ly$\gamma$.
It is thus likely that the absorbing cloud at $z_{\rm abs}$~=~2.2215 
completely covers the continuum source and partially covers the BLR.
%
%

This system does not show absorption due to any detectable
heavy element transitions
either in the optical data (Outram et al 1998; Savaglio, 1998)
or in the HST data. 
If partial coverage is a signature of physical association between the
absorbing gas and the AGN, then the lack of metal lines in this system 
could suggest that there are large inhomogeneities in the chemical 
enrichment of
the gas physically associated with central engines of QSOs.
However, on the contrary, if we presume that the gas associated with the 
central regions of the quasar is more or less uniformly enriched, then this 
system could correspond to very highly ionized gas. The ionization state 
should be such that all observable metal transitions are weak and
undetectable. This condition demands log H~{\sc i}/H to be smaller than -8
(e.g. Hamann 1997)
and log $N$(H)$>$22. Note that such a cloud could be related to the
warm absorbers.

Another possibility is that the gas is extremely metal-poor and is
produced by an intervening cloud with sizes less than the
BLR (i.e. few pc). Not only this is much smaller than the dimensions derived 
for intervening \lya clouds using adjacent lines of sight (e.g.
Petitjean et al. 1998) but also these clouds would have been detected 
by previous surveys far away from the QSO.
\subsection{\zabs = 2.207}
The \civ, \nv and \lya absorption lines produced by this system are
shallow and broad ($\sim$500~km~s$^{-1}$) like a miniaturised Broad
Absorption Line system (BALs). Outram et al. (1998) discuss the \nv and
\lya absorptions from this system. They could not fit the \nv doublet
when assuming 100\% coverage. However, they managed to obtain a
consistent fit after correcting the continuum by subtracting a
Gaussian centered at 3938~\AA~ with FWHM = 670 \kms~ and maximum depth
19 percent of the original continuum level. They used a three-component
model with large velocity dispersions. Savaglio (1998) observed the
\civ absorption doublet from this system. She could fit the doublet
with six components. The \ovi and Ne~{\sc viii} doublets are detected
in the G430M and E230M spectra respectively.  Note that the \ovi
doublet is observed at slightly lower resolution than \nv and \civ and
the Ne~{\sc viii} doublet is found in a low S/N region blueward the
Lyman limit of the moderately thick system at \zabs = 1.87. Finally,
strong O~{\sc v}$\lambda$629 is seen at $\sim$2020~\AA~ as expected and
possibly Mg~{\sc x}$\lambda$624 at $\lambda$2000~\AA.

%
\begin{figure}
\psfig{figure=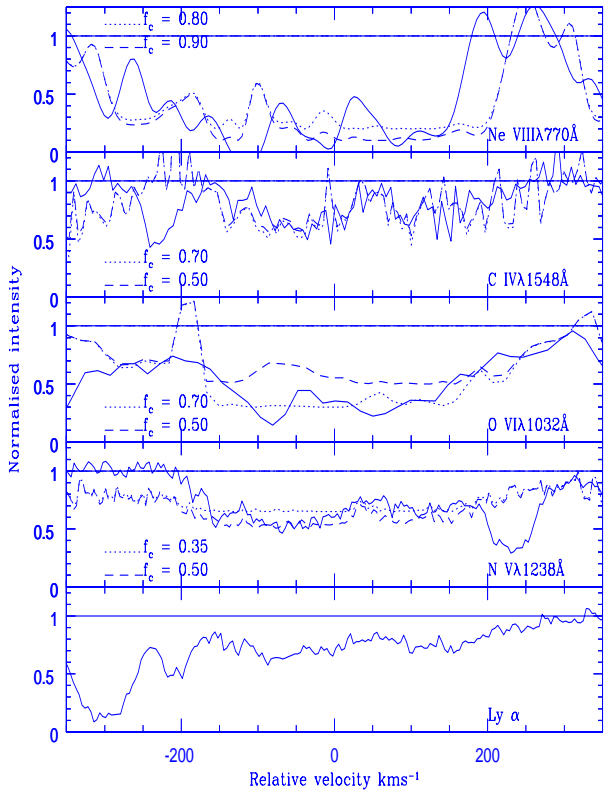,height=11.5cm,width=8.5cm,angle=0}
\caption[]{Analysis of partial coverage in the \zabs = 2.207 system.
The observed \lya profile is plotted in the bottom panel. The solid curves in
the other panels are the observed profiles of the stronger line of the
doublets. The dotted and short-dashed lines are the
predicted velocity profiles computed from the profile of the second 
transition of the doublet and assuming covering factors
$f_{\rm c}$~=~0.35 (dotted line) and 0.50 (dashed line).}
\label{f220}
\end{figure}
%
%

The stronger line of the \civ, \nv, \ovi and Ne~{\sc viii} doublets, 
together with the \lya line, are plotted on Fig.~\ref{f220} 
on a velocity scale. The dotted and dashed lines are the predicted velocity
profiles computed from the profile of the second transition of the doublets 
for different values of the
covering factor. Here again we assume identical covering factors for both
transitions. 
It can be seen that in order to reproduce
the residual intensities of \nv we need a covering factor of the order of
0.35 at $v$~$\sim$~+150 \kms~ and 0.50 at $v$~$\sim$~--100 \kms.
The difference between the two models with $f_{\rm c}$~=~0.35 and 0.50, 
is, at these places, of the order or larger than
0.1 (10\% of the normalized continuum; see Fig.~\ref{f220}) when the 
rms deviation in the spectrum is $\sim$0.04. 

The \ovi profiles are consistent with a covering factor
$\sim 0.7$ that cannot be reconciled with the values found for \nv.
Note that the \ovi profiles  could be
affected by the relatively poor spectral resolution and the derived
covering factors should be considered as lower limits. The maximum
error in the covering factor for the \ovi lines is about 0.1, computed from 
the error in the residual intensity.
The S/N ratio over the \civ doublet is not good enough and consistent 
residual intensities are obtained for a wide range of covering
factors. 
The covering factor
required for the Ne~{\sc viii} lines is in the range 0.8-1.0 with an
error per pixel of 0.15. The covering factor derived
from the Ne~{\sc viii} lines is therefore significantly larger than the
one derived from
the \nv doublet. 
There seems to be an anti--correlation between the covering factor and
the ionization state or the wavelength. This again is consistent with
clouds partially covering the BLR while covering most of the
continuum emission region.

Another interesting observation concerns the doubly-ionized species.
It can be seen on Fig.~\ref{transi} that 
O~{\sc iii}$\lambda$832 is certainly blended with other lines;
that there may be a shallow absorption
at the expected position of C~{\sc iii}$\lambda$977 although most of
it should be Ly$\gamma$; and that there 
is a strong absorption at the expected position of N~{\sc iii}$\lambda$989.  
This line cannot be \lyb at $z_{\rm abs}$~=~2.093 as there is no
corresponding \lya line. We cannot reject the hypothesis that this is
an intervening \lya line as the \lyb range is of very poor S/N ratio. 
Interestingly enough, there is an absorption feature at the expected
position of N~{\sc iii}$\lambda\lambda$684,685
(see Fig.~\ref{transi2}).
Although such a strong N~{\sc iii} absorption would be very 
surprizing (see next Section and Fig.~\ref{models}),
the presence of doubly-ionized species cannot be ruled out.
The corresponding absorptions from triply ionized species, 
although weak and noisy, could be present (see in particular 
O~{\sc iv}$\lambda$787 and N~{\sc iv}$\lambda$765).
If true, this would imply that the medium has two phases of low and
high-ionization (see next Section).

Since the possible strong N~{\sc iii}$\lambda$989 line goes to zero,
suggesting complete coverage; we note that
there is a tendency for the absorption lines redshifted on top of the 
emission lines (here \civ and \nv) to have covering
factors smaller than lines redshifted in parts of the
spectrum free from emission-lines (Ne~{\sc viii} and N~{\sc iii}).
This further supports the conclusion that 
the gas covers the continuum emitting region but only 
part of the BLR.

\subsection{\zabs = 2.198 }
\lya and \nv absorptions produced by this system are detected by Outram
et al. (1998) who already noted the incomplete coverage of
the background source. They conjectured that the absorbing cloud is larger
than the continuum emitting region and smaller than the BLR. Savaglio
(1998) has noted that single as well as two component fits to the \civ
doublet result in a very poor fit, again suggesting partial
coverage. \ovi and Ne~{\sc viii} doublets are detected in the HST
spectra.
\begin{figure}
\psfig{figure=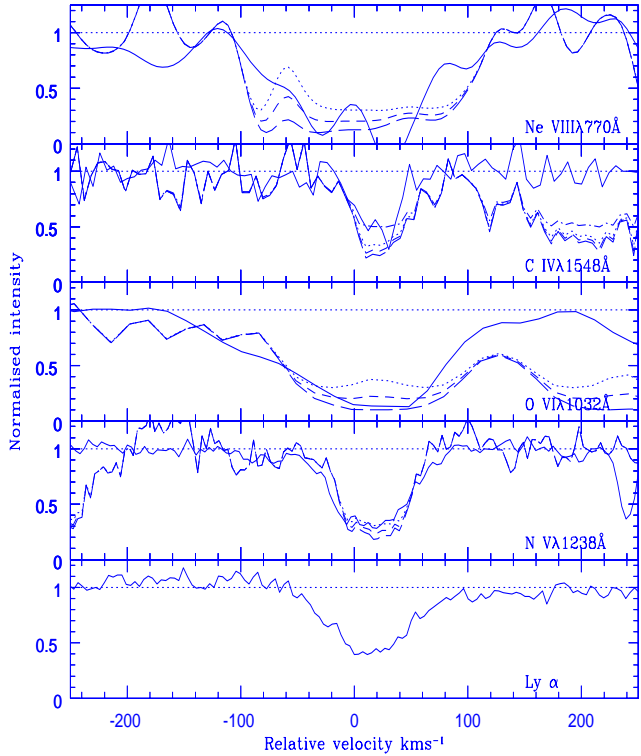,height=11.cm,width=9.cm,angle=0}
\caption[]{Analysis of partial coverage in the \zabs = 2.198 system.
The observed \lya profile is plotted in the lowest panel. The solid curves in
other panels are the observed profiles of the strongest transition of the
doublet. The dotted, short-dashed and long-dashed lines are the
predicted velocity profiles computed using the profile of the weakest member 
of the doublets and assuming covering factor of $f_c$ = 0.7, 0.8 and 0.9 
respectively.}
\label{f219}
\end{figure}
The stronger lines of the doublets together with \lya are plotted on
Fig.~\ref{f219} on a velocity scale. The dotted and dashed lines are the
predicted velocity profiles computed from the second transition of doublets 
assuming different values of the covering factor (dotted, short-dashed and
long-dashed lines are for $f_c$ = 0.7, 0.8 and 0.9 respectively). Here
again we assume identical covering factors for both transitions.
The \nvb line of
this system is blended with \nva of the system at 2.208.  We have
subtracted the contribution due to the \nva line before doing the
analysis. The covering factor required to fit the \nv doublet is
$\sim0.7\pm0.04$. However the \ovi profiles require values
larger than 0.85$\pm0.04$. 
It is quite likely that the Ne~{\sc viii}$\lambda$780 is
blended with some other line in the blue wing. Also the Ne~{\sc
viii}$\lambda$770 profile is very noisy and is consistent with a wide
range of covering factors. It is interesting to note that inspite of
the poor signal to noise ratio, the \civ profiles clearly suggest that the
covering factor is less than 0.7 and we have obtained a consistent
fit for $f_c$ = 0.5. Thus like the \zabs = 2.207 system, this system also
shows different covering factor for different transitions; 
the covering factor beeing lower for \civ than for \nv.

\section {Physical conditions in the \zabs~=~2.198 system} 
%
%
%

In this section we study the physical conditions in the \zabs = 2.198
system in greater detail. The column density per unit
velocity interval at any velocity, $v$, with respect to the centroid of
the line is given by,
\begin{equation}
N(v)~=~{(m_ec/\pi e^2)\over f \lambda}\tau(v),  
\end{equation}
and from Eq.~(1) the optical depth $\tau(v)$ is given by,
\begin{equation}
\tau(v)~=~-~ln~\bigg({R(v)-1+f_c\over f_c}\bigg).
\end{equation}
Assuming equal covering factor for the two absorption lines of a 
doublet, $f_{c1}=f_{c2}=f_{c}$,  we
obtain $f_c(v)$  for each doublet using Eq.~(2). We then derive the
column density per unit velocity interval from Eqs.~(3) and (4).
The total column density, $N$, is obtained by integrating $N(v)$ over
the velocity interval covered by the absorption line profile. 
For various species we give in Table~2 the
absorption parameters ($N$ and $b$) obtained from Voigt profile fitting
(column \#3 and \#4 respectively) and the column density obtained by
integration of the column density per unit velocity interval over the
velocity profile (column \#7). 
The velocity range (column \#5), and the mean covering
factor estimated over this range (column \#6) 
are also given. In the case of single lines we use
a conservative lower limit of 0.7 for the covering factor. Note that
the column density obtained with the Voigt profile fits are lower
limits as corrections are not incorporated to take into account 
partial coverage. 

\lyb from this system is weak and we could not derive any bound on the 
covering factor from the \lya line only.
\ciii is not detected and we derive a two sigma upper limit from the
continuum rms at the expected position of the line. Voigt profile fits
to the \civ lines are taken from Savaglio (1998). As the estimated
covering factor is low, the column density derived by fitting the
profile is less by a factor of 5 compared to the estimated column
density using the column density per unit velocity interval.
%
%
N~{\sc iii} and Si~{\sc iii} are not detected whereas a line is 
present at the expected position of N~{\sc iv}$\lambda$765. 
%
%
This line clearly shows 
two components and the blue
component is clearly absent in the \nv profile. The further absence of N~{\sc
iii} suggests that this component is not real. Thus we have fitted the
N~{\sc iv} blue component as an intervening \lya line and used only the
column density obtained from the red component in our analysis. 
Although the resolution of the spectrum is not very high ($\sim$ 50 \kms),
we have fitted the \ovi lines using
three components. 
There is a line at the expected position of O~{\sc iv}$\lambda$787.
However this line could possibly be Ly$\gamma$ from the system at 
\zabs = 2.5907. By carefully fitting the \lya, \lyb and Ly$\delta$ lines 
in this system, we removed the contribution of the Ly$\gamma$ and 
fit the residual as O~{\sc iv}$\lambda$787.
As discussed before, the estimate of column densities
can be somewhat uncertain due to limited S/N ratio (especially for the 
Ne~{\sc viii} lines) and possible contamination by intervening lines.

\begin{table*}
\begin{tabular}{lccccrr}
\multicolumn{7}{l}{{\bf Table 2.} Parameters for the associated system 
at \zabs=2.198 }\\ 
\hline
\multicolumn{1}{c}{Ion}&\multicolumn{1}{c}{$z$}&
\multicolumn{1}{c}{log $N$}&\multicolumn{1}{c}{$b$}&
\multicolumn{1}{c}{$v$}&\multicolumn{1}{c}{$f_{\rm c}$}&
\multicolumn{1}{c}{log $N$}\\
\multicolumn{1}{c}{ }&\multicolumn{1}{c}{ }&
\multicolumn{1}{c}{cm$^{-2}$}&\multicolumn{1}{c}{(\kms)}&
\multicolumn{1}{c}{(\kms)}&\multicolumn{1}{c}{ }&
\multicolumn{1}{c}{cm$^{-2}$}\\
\hline\\
\\
H~{\sc i} &  2.1982&13.74$\pm$0.01&43.99$\pm$1.01&-47, 53& $>$0.7&$<$13.95\\
C~{\sc iii}&  .....&  .....  & ..... & .....  &.....&$<$13.40\\
C~{\sc iv}&  2.1982&13.77$\pm$0.05&19.80$\pm$2.40&-27, 50& $\sim$0.5&14.48\\
N~{\sc iv}&  2.1980&13.73$\pm$0.04&30.43$\pm$1.35&       & $>$0.7&$<$14.34\\
N~{\sc v} &  2.1972&13.12$\pm$0.05&32.51$\pm$4.06&\\
          &  2.1982&14.57$\pm$0.03&31.46$\pm$0.58&-27, 53& $\sim$0.7&14.55\\
O~{\sc iv}&  2.1979&14.43$\pm$0.06&49.33$\pm$2.74&&&\\
          &  2.1987&13.94$\pm$0.08&37.70$\pm$4.53&-47,  53& $>$0.7&$<$14.62\\
O~{\sc vi}&  2.1976&14.39$\pm$0.16&$^a$&&&\\
          &  2.1981&14.87$\pm$0.24&$^a$&&&\\
	  &  2.1987&14.63$\pm$0.14&$^a$&-46, 100&$\sim$0.8&15.38\\
Ne~{\sc viii}&2.1976&14.63$\pm$0.27&39.11$\pm$9.70&&&\\
             &2.1981&14.79$\pm$0.10&25.39$\pm$3.85&&&\\
	     &2.1987&14.68$\pm$0.09&32.33$\pm$3.20&-27, 53& $\sim$0.7&15.10\\
 & & & &\\
\hline 
\multicolumn{7}{l}{$^a$ Same as for Ne~{\sc viii}}\\       
\end{tabular}
\end{table*}
\begin{figure}
\centerline{\vbox{
\psfig{figure=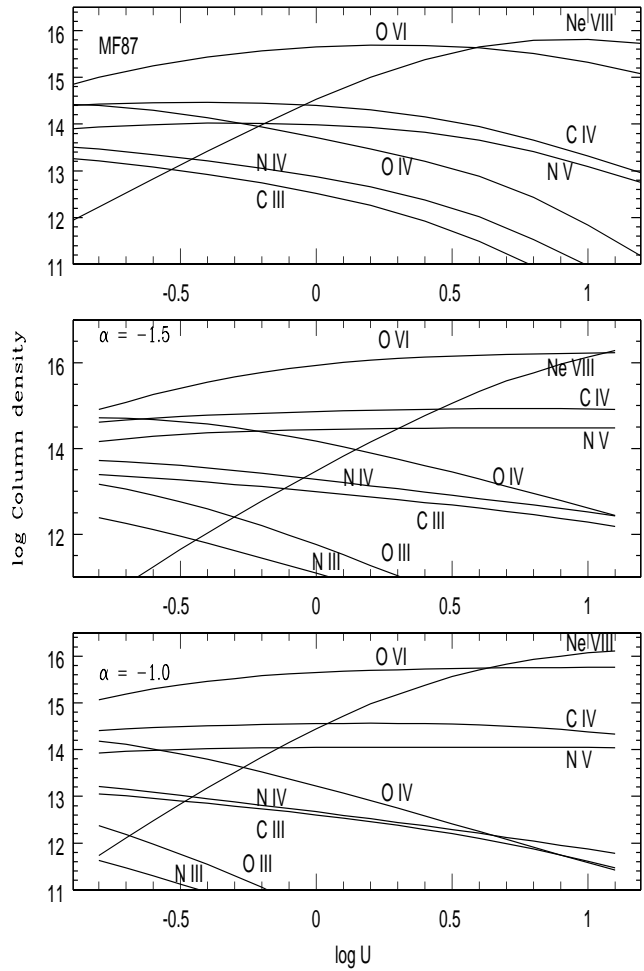,height=14.cm,width=9.cm,angle=0}
}}
\caption[]{
The logarithm of column densities for species indicated next to each curve 
is given versus the logarithm of the ionization parameter for ionizing 
spectra as given by Mathews \& Ferland (1987; top panel); or power-laws
with index $\alpha$~=~-1.5 (middle panel) and -1 (bottom panel).
}
\label{models}
\end{figure}

The derived column densities for H~{\sc i}, \civ, \nv and \ovi are
characteristic of associated systems (Hamann 1997).
We run photo-ionization
models using Cloudy (Ferland 1996) to study the ionization structure
of a plane parallel cloud with neutral hydrogen column density
10$^{14}$~cm$^{-2}$, solar metallicities, and illuminated by ionizing 
radiation fields with different spectra. Although a one-zone model is
questionable in such medium, we believe that this is a
reasonable representation of the absorbing cloud 
producing at least C~{\sc iv}, N~{\sc v} and H~{\sc i}
(see below for O~{\sc vi} and Ne~{\sc viii})
because the kinematics of the lines 
are very similar (see Figs.~1 and 5) and the
partial coverages discussed in the previous sections 
indicate small sizes for the cloud. \par\noindent

Results for a typical AGN spectrum 
given by Mathews \& Ferland (1987) and two power-law spectra are given in
Fig.~\ref{models}.
In the framework of these models,
it is apparent that the column densities are reproduced easily.
The $\alpha$~=~-1 power-law ionizing spectrum is favored as it
minimizes the $N$(C~{\sc iv})/$N$(O~{\sc vi}) ratio.
Although metallicities cannot be much less than solar
(because of \civ and \nv), it would be difficult to argue for metallicities 
much larger than solar (especially for oxygen).
However, contrary to what is observed, the predicted \civ
column density is always larger than that of \nv. This suggests that
nitrogen is over-abundant compared to carbon.
%
%
%
%
It is interesting to note that
the \nv lines of the \zabs = 2.207 system is also stronger than the lines 
of \civ suggesting a similar abundance pattern. Indeed, using \nv/\civ
emission line ratios, Hamann \& Ferland (1992)
have shown that nitrogen is over-abundant by a factor of $\sim 2-9$ in
high-redshift QSOs ($z\ge 2.0$). They suggested rapid star-formation
models to boost the nitrogen abundance through enhanced secondary
production in massive stars. Korista et al. (1996) have  also noticed
overabundance of nitrogen with respect to carbon and oxygen in
the well-studied BAL system in Q~0226-1024. They could obtain a much better
fit of the lines after taking into account the abundance pattern due to
rapid star-formation. The overabundance of nitrogen with
respect to oxygen and/or carbon, does not seem to be seen in every
associated
system however. Indeed, we have good data for the associated systems in 
Q~0207--003 and Q~0138--381 showing partial coverage. Although 
solar metallicities are needed to explain the line ratios, there is no
indication of enhanced nitrogen abundance.


 


%
From the observed \nv to N~{\sc iv} column density ratio, it can be seen that
log~$U<$ --0.5 (the ionization parameter $U$ is the ratio of the
ionizing photons density to the total hydrogen density). 
However such a value for the ionization parameter implies that 
the gas producing \nv and N~{\sc iv} can not account for the observed value 
of the Ne~{\sc viii} column density without an unrealistically large neon 
abundance.
Another and more likely explanation is that there are two  distinct regions
with different ionization parameters and that
Ne~{\sc viii} predominantly originates from the more highly ionized region. 
This is supported by the fact that the Ne~{\sc viii} absorption is spread
over a larger velocity range than the \nv and \ovi absorptions.  
We can check that, even in that case, C~{\sc iii}$\lambda$977 would not be
detectable. Indeed, the expected C~{\sc iii} column density is of the order
of 10$^{13}$~cm$^{-2}$ (see Fig.~\ref{models}), thus 
$w_{\rm obs}$(C~{\sc iii}$\lambda$977)~$\sim$~0.2~\AA~
which is about twice below the detection limit at $\lambda$~$\sim$~3100~\AA.


Since the low-ionization region by itself can account for 
the observed H~{\sc i} and \ovi column densities, the 
high-ionization region should have $N$(Ne~{\sc viii}) much larger than 
$N$(\ovi) which means log~$U$ larger than 0.5 and most probably 1. 

If we suppose that this component is similar to warm absorbers 
detected by O~{\sc vii} and O~{\sc viii} edges in the X-rays,
then the condition that the optical depth
of these edges are larger than 0.1 implies log~$N$(O~{\sc vii})~$>$~17.55 
and log~$N$(O~{\sc viii})~$>$~18.00. 
Photoionization models with solar abundances, fail to produce such 
high O~{\sc vii} and O~{\sc viii} column densities for H~{\sc i}
column densities in the range 10$^{13}$ to 10$^{14}$~cm$^{-2}$. 
Indeed, the ratios NeVIII/Ne, OVII/O and OVIII/O are all maximized 
over the range of ionization parameters $U$~$\sim$~10--100 
(see Hamann et al. 1995) where log~HI/H~$\sim$~--6. For 
log~$N$(H~{\sc i})~=~14, this corresponds to log~$N$(H)~$\sim$~20 and,
assuming solar abundances, implies that log~$N$(O~{\sc vii}) and 
log~$N$(O~{\sc viii})
are less than~17. Thus, it is most likely that in QSO J2233-606,
the region producing the Ne~{\sc viii} absorption can not be a
warm absorber. It is thus of first importance to study the 
intrinsic spectral energy distribution of J2233-606 especially in the
X-rays.

%

%

\section{Conclusion}

We have studied the associated absorption systems 
($z_{\rm abs}$~$\sim$~$z_{\rm em}$) in QSO J2233-606, 
seen over the redshift range 2.198--2.2215, corresponding to outflow 
velocities relative to the quasar emission redshift ($z_{\rm em}$~=~2.252
from Mg~{\sc ii}) of 2800--5000~km~s$^{-1}$ which is modest
compared to usual associated or BAL outflows.

We have shown that the Ly$\alpha$ line at \zabs = 2.2215 is saturated
but has non-zero residual intensity. The covering factor of the gas is of
the order of 0.7. This absorption system is unusual in the sense that
there is no detectable metal lines. This could reveal chemical
inhomogeneities in the gas or, and more likely, this absorption could
correspond to very highly ionized gas from which no absorption due
to heavier element can be
detected. If the latter is true, the ionization factor log~HI/H could be
smaller than --8 and the total column density in the cloud larger than
log~$N$(H)~=~22. Such a cloud could be related to the warm absorbers. 

Over the redshift range 2.198--2.21, conspicuous Ne~{\sc viii}, O~{\sc vi},
C~{\sc iv} and N~{\sc v} absorptions are seen whereas the H~{\sc i} 
absorption is weak. From analysing these absorptions we conclude that (i) a 
two-phase medium of low and
high-ionization respectively is required to explain the complete set of column
densities; (ii) the abundances are close to the solar value but 
nitrogen is enhanced with respect to carbon; (iii) although difficult to 
ascertain there is a tendency for the 
absorption lines redshifted on top of the emission lines to have covering
factors smaller than lines redshifted in parts of the
spectrum free from emission-lines; this suggests that 
the gas covers the continuum emitting region but only part of the 
BLR.
The component at $z_{\rm abs}$~=~2.21 (+950~km~s$^{-1}$ on Fig.~\ref{transi}) 
is remarkable as, though conspicuous in Ne~{\sc viii} and 
O~{\sc vi}, it is undetected in H~{\sc i}, \civ and \nv. The column densities 
derived for this subcomponent by Voigt-profile fitting are
log~$N$(Ne~{\sc viii})~=~14.57$\pm$0.10; log~$N$(O~{\sc vi})~=~14.05$\pm$0.06 
and log~$N$(H~{\sc i})~$<$~13.30, consistent with what is expected
from a high-ionization zone. 


\acknowledgements{We would like to thank all astronomers who have provided
the data for this project and especially the HST HDF-S STIS team lead
by H. Ferguson. We thank the referee, Fred Hamann, for
a careful reading of the manuscript.}

\end{document}